\documentclass[aps,pre,twocolumn,showpacs]{article}
\usepackage{bm}
\usepackage{amsmath}
\usepackage{amssymb}
\usepackage{amsfonts}
\usepackage{graphicx}

\begin{document}

\author{M. Stepanova(1), E. E. Antonova(2), J.M. Bosqued(3) 
\\ 
\\
\\{\it (1) Departamento de F\'isica,
Facultad de Ciencia, Universidad de Santiago de Chile}
\\{\it Casilla 307, Correo 2, Santiago, Chile, mstepano@lauca.usach.cl}
\\
\\  {\it (2) Skobeltsyn Institute of Nuclear Physics, Moscow State University, }
\\{\it Vorobievi Gori 119992, Moscow, Russia, antonova@orearm.msk.ru}
\\
\\ {\it (3) Centre d’Etude Spatiale des Rayonnements, }
\\{\it CNRS/UPS, BP 4346, 31028, Toulouse, France (bosqued@cers.fr)} }

\title{Radial distribution of the inner magnetosphere plasma pressure using low-altitude satellite data during geomagnetic storm: the March 1-8, 1982 Event}

\maketitle
\begin{abstract}
Plasma pressure distribution in the inner magnetosphere is one of the key parameters for understanding the main magnetospheric processes including geomagnetic storms and substorms.
 However, the pressure profiles obtained from in-situ particle measurements by the high-altitude satellites inside the plasma sheet and other regions of the magnetosphere do not allow 
tracking the pressure variations related to the storms and substorms, because a time interval needed to do this generally exceeds the characteristic times of them. On contrary, fast 
movement of low-altitude satellites makes it possible to retrieve quasi-instantaneous radial or azimuthal profiles of plasma pressure along the satellite trajectory, using the precipitating 
particle flux data in the regions of isotropic plasma pressure. For this study, we used the low-altitude polar-orbiting Aureol-3 satellite data for plasma pressure estimation, and the IGRF, 
Tsyganenko 2001 and Tsyganenko 2004 storm time geomagnetic field models for the pressure mapping into the equatorial plane, and for the evaluation the corresponding volume of the
magnetic flux tube. It was found that during quiet geomagnetic condition the radial pressure profiles obtained coincide with the profiles, obtained previously from the high-altitude measurements. 
On the contrary, the plasma pressure profiles change significantly during the development of storms and substorms. Nevertheless, three geomagnetic field models gave significantly different 
geomagnetic field profiles, those points out the necessity to develop a magnetically self-consistent model for description of the inner magnetosphere geomagnetic field. However, the common 
features observed for all models are: during geomagnetic storm the plasma pressure profiles became sharper; the position of the maximum of plasma pressure corresponds to expected one for 
given Dst minimum; the maximum value of inner magnetosphere static pressure correlates with the solar wind dynamic pressure. Increase in the plasma pressure profiles indicates the possibility 
to consider the interchange instability as one of important factors for the development of the main phase of geomagnetic storm. 
\end{abstract}

\section{Introduction}
As it is well known, strong geomagnetic storms take place when the interplanetary magnetic field (IMF) has a prolonged southward orientation. During storms a powerful ring current is developed, the intensity of which is commonly evaluated by Dst index, introduced by M. Sugiura (1964). Contribution of the ring current into the Dst variation is determined by the total inner magnetospheric energy content of the trapped particle population, which depends on particle supply, energization and losses. The inner magnetospheric energy content is generally evaluated using the Dessler-Parker-Sckopke-Tveskoy relation (Dessler and Parker, 1959; Sckopke 1966; Parker, 1996; Tverskoy, 1997). Recently, this relationship was modified by Antonova (2002) and Liemohn (2003), taking into consideration that the value of plasma pressure at the external boundary of the ring current region is not equal to zero. However, during last decade it was proposed that the tail current makes significant contribution into Dst variation. This statement
was based on the fact that the measured by Dst storms strength or its rate of strengthening can be decreased during substorm expansion phase (Iyemori and Rao, 1996). It was also found that, on average, during the geomagnetic storm main phase, Dst decreases more slowly after a substorm expansion phase starts than before, and that on average, during the storm's recovery phase, it increases more rapidly (Iyemori and Rao, 1996, McPherron, 1997). 
Nevertheless, the traditional interpretation of Dst as a measure of ring current intensity can be reestablished by modifying the topology of high latitude current systems (Antonova, 2004). As it is well known, energetic particles move around the Earth near equator in the midnight sector and at high latitudes near noon in the region of quasi-trapping. Antonova and Ganushkina (2000) suggested that the existence of daytime high-latitude field minima leads to the splitting of the high latitude ring current into two branches. The suggested current system was named the cut-ring current (CRC) system. It was shown that the transverse currents, created by the gradients of the plasma pressure at L more than 7 till the magnetopause, are concentrated far from the equatorial plane in the daytime sector of the magnetosphere, being a high latitude continuation of ordinary ring current, distorted due to daytime compression of the magnetosphere.  This makes it possible to turn back to the traditional interpretation of Dst as a result of ring current development, taking into account energy content of the region of quasi-trapping in the ring current energy budget that includes the nighttime part of the plasma sheet till aproximately 12 Re (Radius of the Earth) during quiet conditions and its daytime continuation. 
The results of Lyons et al. (2003) made a significant contribution in understanding of processes leading to the decrease in the absolute value of Dst or in the rate of strengthening of geomagnetic storms. In particular, they have shown that substorm injections lead to the decrease in the plasma pressure at the geocentric distances 10-13 Re. Pressure reductions after the substorm expansion phase onset at the geostationary orbit were also observed earlier by Roux (1985) and Kozelova et al. (1986). So, we can assume that plasma transport by the large-scale electric field during substorm growth phase leads to the increase of Dst. Otherwise, losses of particles after the substorm expansion phase onset (probably in the tailward direction) explain the decrease of ?Dst? or the decrease of the rate of strengthening of the geomagnetic storm. So, the correct consideration of pressure balance in the inner magnetosphere is very important for understanding of the ring current dynamics during geomagnetic storms. However, for the correct estimation of this balance it is necessary to include the ionospheric sources of ions. In particular,  it was found that the ions of oxigen provide more than aproximately 40 percent of particle energy density during the main phase of great storms and aproximately 20 percent during small to moderate storms (Daglis, 1997, Pulkkinen et al., 2001).
Distribution of plasma pressure in the magnetosphere of the Earth has been studied extensively during last decades. In particular, Lui et al. (1987), Lui and Hamilton (1992) obtained profiles of plasma pressure using in situ particle measurements onboard the high-altitude AMPTE/CCE satellite. Later, De Michelis et al. (1999) also obtained bi-dimensional distribution of plasma pressure in the equatorial plane, using the data of the same satellite. Lui (2003) studied the magnetic local time (azimuthal) asymmetry in the inner magnetosphere plasma pressure. Wing and Newell (1998, 2000) reproduced statistical bi-dimensional distribution of plasma pressure in the equatorial plane at aproxiamtely 10 Re using the low-altitude DMSP series satellites and determined the corresponding Region 1 field-aligned current distribution. Stepanova et al., (2002) modified the technique proposed by Wing and Newell (1998, 2000), considering the presence of a field-aligned potential drop in the auroral geomagnetic field lines, and obtained footprints of plasma sheet pressure profiles in the equatorial plane for very short time intervals (minutes) for quiet geomagnetic conditions and during different substorm phases. This is impossible to do using only the high-altitude satellite measurements. Stepanova et al., (2004a) evaluated the quasi-instantaneous value of azimuthal plasma pressure gradient and corresponding values of field-aligned currents, using oblique passages of the low-altitude Aureol-3 satellite, that agreed with typical values of Region 1 and 2 field-aligned currents. 
Study of plasma pressure profile during magnetic storms is constrained because of appearance of energetic particle fluxes, mainly relativistic electrons that can cause significant distortions in the instrument functionality and even the satellite devices. The main contribution to the storm time ring current plasma pressure is introduced by ions with energies 100 keV, approximately, (see Lyons and Williams, 1982; Williams, 1983). However existence of CRC requires analysis of particle fluxes with energies of 10 keV, approximately, forming part of particle population responsible for part of Dst variation.
In this paper we use the methodology developed by Stepanova et al., (2002) for studying the evolution of the radial pressure profiles, obtained by the Aureol-3 satellite before, during and after the March 1-8, 1982 geomagnetic storm and establish the upper limit of the increase in steepness of low energy the part of magnetospheric plasma pressure profile connected to the ion fluxes with energies 1-22 keV. We try to show that this limit is determined by the development of interchange (flute) instability.  

\section{Evolution of the radial plasma pressure profiles during March 1-8, 1982 geomagnetic storm}

Despite significant efforts concentrated on the determination of distribution of the plasma pressure in the inner magnetosphere, the evolution of this distribution during geomagnetic storms is not established. As was mentioned before, we propose to obtain nearly-instantaneous radial plasma pressure profiles at the external part of the ring current using low-altitude satellite data. The main difficulties of this method consist in the necessity to use some geomagnetic field model for mapping of the measured plasma pressure profile into the equatorial plane and possible anisotropy of particle fluxes. High apogee satellites obtain the plasma pressure in situ, nevertheless they move comparatively slowly, and it takes hours to obtain a radial pressure profile. This time is too large to guarantee what the profile obtained remains unchanged, and this approach can not be used for geomagnetic storm studies. Use data of low orbiting satellites makes it possible to restore plasma pressure profiles only in the regions of isotropic pressure. Therefore it can give only partial information about the plasma pressure profile. Nevertheless, this kind of information gives the possibility to evaluate the lower limit of the value of plasma pressure in the equatorial plane and steepness of radial plasma pressure profile.

\onecolumn

\begin{figure}[t]
\centering \vspace{-0.0cm} \hspace{-0.3 cm}
\includegraphics[width=.8 \textwidth]{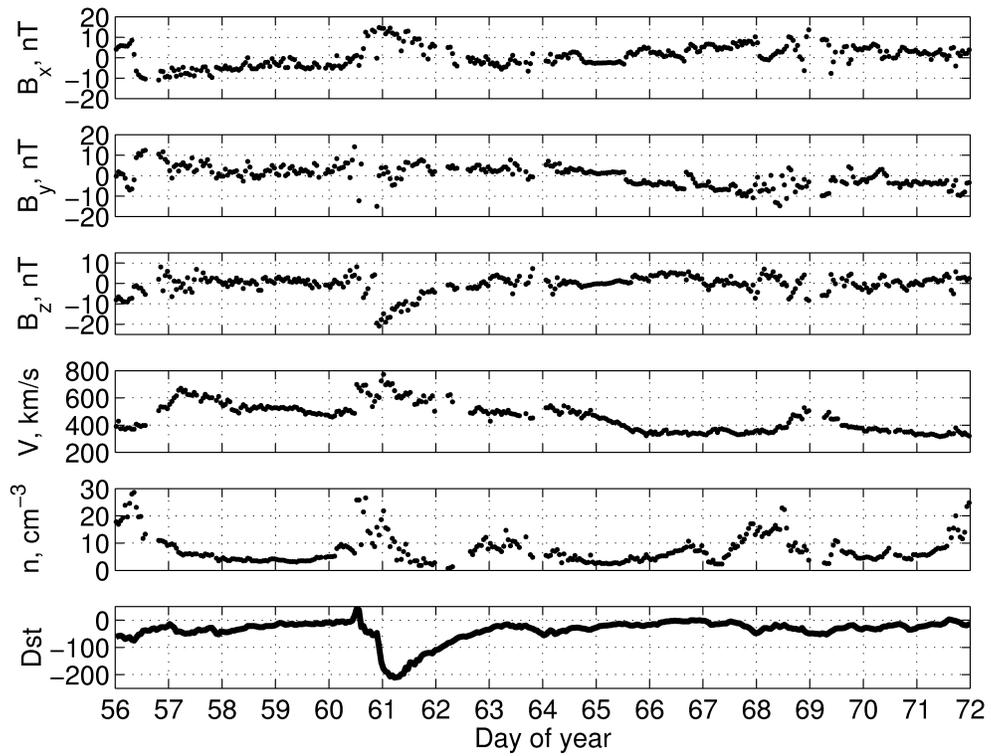}
\caption{Variations of the solar wind and Dst index between February 25 and March 12, 1982. From top to bottom: IMF $B_{x}$, $B_{y}$, $B_{z}$, in GSE, the solar wind velocity and number density, and Dst index. Data were downloaded from the OMNI 2 data base (http://omniweb.gsfc.nasa.gov/)}
\label{one}
\end{figure}

\begin{figure}
\centering \vspace{-0.0cm} \hspace{-0.3 cm}
\includegraphics[width=.8 \textwidth]{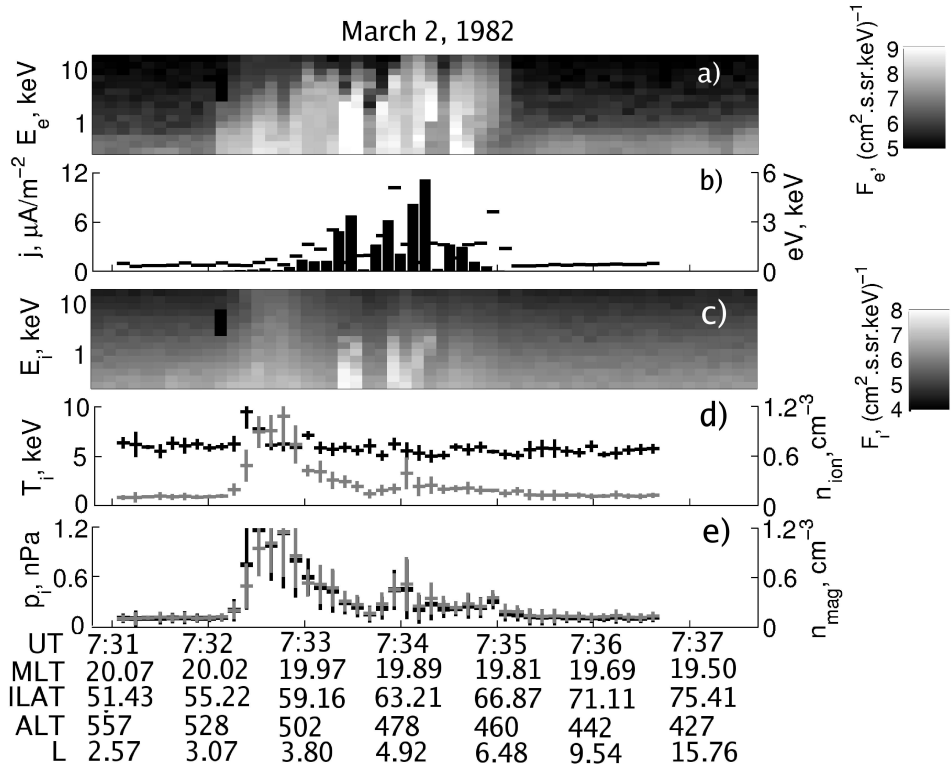}
\caption{The electron precipitating fluxes (a), the field-aligned current densities and field-aligned potential drops (b), the ion precipitating fluxes (c), the ion concentration at the ionospheric altitudes and the ion temperature (d), the ion concentration in the magnetosphere and the plasma pressure (e), obtained during the March 2, 1982 auroral oval crossing.}
\label{one}
\end{figure}

\twocolumn
Here we use the same methodology as in (Stepanova et al., 2002) to study the low energy plasma pressure profile during the March 1-8, 1982 intense geomagnetic storm (minimum Dst 211 nT). The interplanetary magnetic field components in GSE and the solar wind velocity and number density and Dst index variations are shown in Figure 1. Pressure profiles were obtained taking into consideration the influence of field-aligned potential drop on the ion precipitations. This was carried out simultaneously with the analysis of the meridional distribution of the electron precipitating flux using the polar orbiting (perigee: 410 km, apogee: 2000 km, inclination: 82.5), three-axis-stabilized Aureol-3 spacecraft. Particle data with a time resolution of 1.6-3.6 s were provided by the SPECTRO instruments within the 0.02-22 keV energy range (Bosqued et al., 1986). 

We analyzed 13 radial profiles of plasma pressure obtained during auroral oval crossing in the sector between 19 and 23 hours of magnetic local time (MLT) before, during and after the March 1-8, 1982 geomagnetic storm. The procedure of low energy plasma pressure profile retrieval is illustrated in Figure 2. In this Figure we can see the electron precipitating fluxes (a), the field-aligned current densities and field-aligned potential drops (b), the ion precipitating fluxes (c), the ion concentration at the ionospheric altitudes and the ion temperature (d), the ion concentration in the magnetosphere and the low energy plasma pressure (e), obtained during the March 2, 1982 auroral oval crossing at 20 MLT, approximately.

This procedure is described in detail in (Stepanova et al., 2002, 2004ab). It is based on the assumption that the ion anisotropy in the equatorial plane is low in the range between 22 and 02 MLT (De Michelis et al., 1999), and the ion distribution functions are nearly Maxwellian. As reported by Antonova and Tverskoy (1975), the field-aligned potential drop decreases the ion concentration at the ionospheric altitude and does not affect the ion temperature. In this case the ion particle flux measured by a low-altitude satellite 
is $I\left( E\right) =\left[ n_{ion}/\left( 2^{1/2}m_{i}\pi^{3/2T_{i}^{3/2}}\right) \right]E\exp\left(-E/T_{i}\right)$. 
Here  is the ion energy, $E$ is the ion energy, $m_{i}$ is the ion mass, and $n_{ion}=n_{mag}\exp \left( -eV/T_{i}\right)$, where ($n_{mag}$) and temperature ($T_{i}$),
obtained from precipitating ion fluxes are shown in Fig. 2(d). Figure 2 (e) shows that low energy plasma pressure has a clear maximum $ $ 1.2  
nPa at 07h32m30s UT. After that the pressure decreases  till $ $ 0.1-0.2 nPa, that satisfies to the total pressure balance conditions in the tail lobes, assuming that in this region the contribution of energetic particles in the plasma pressure is not significant, and that the value of the magnetic field in the lobes is $ $ 20 nT, except a small variation at 07h34m00s UT.

\begin{figure}[t]
\centering \vspace{-0.0cm} \hspace{-0.3 cm}
\includegraphics[width=.5\textwidth]{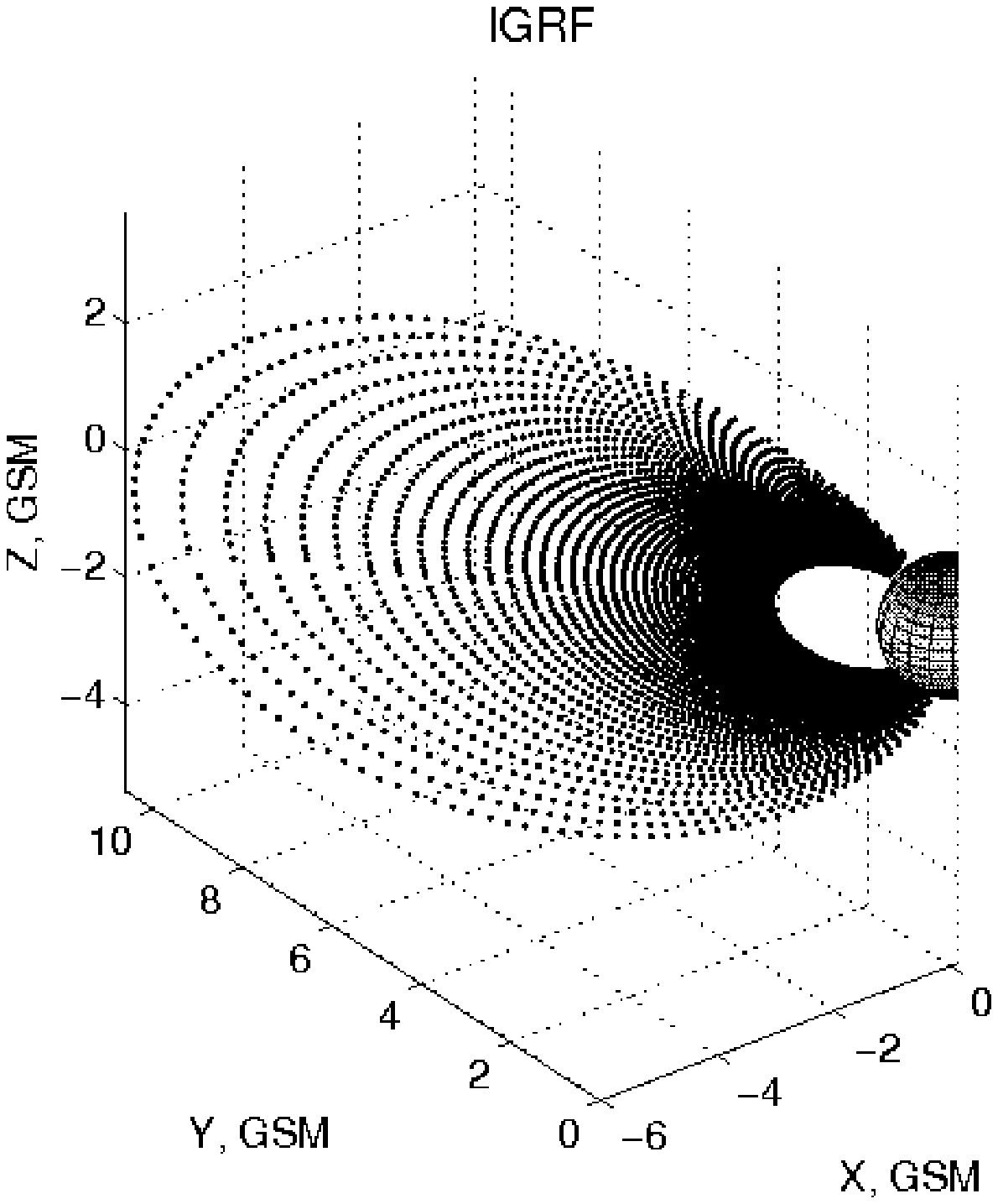}

\label{one}
\end{figure}
\begin{figure}
\centering \vspace{-0.0cm} \hspace{-0.3 cm}
\includegraphics[width=.5\textwidth]{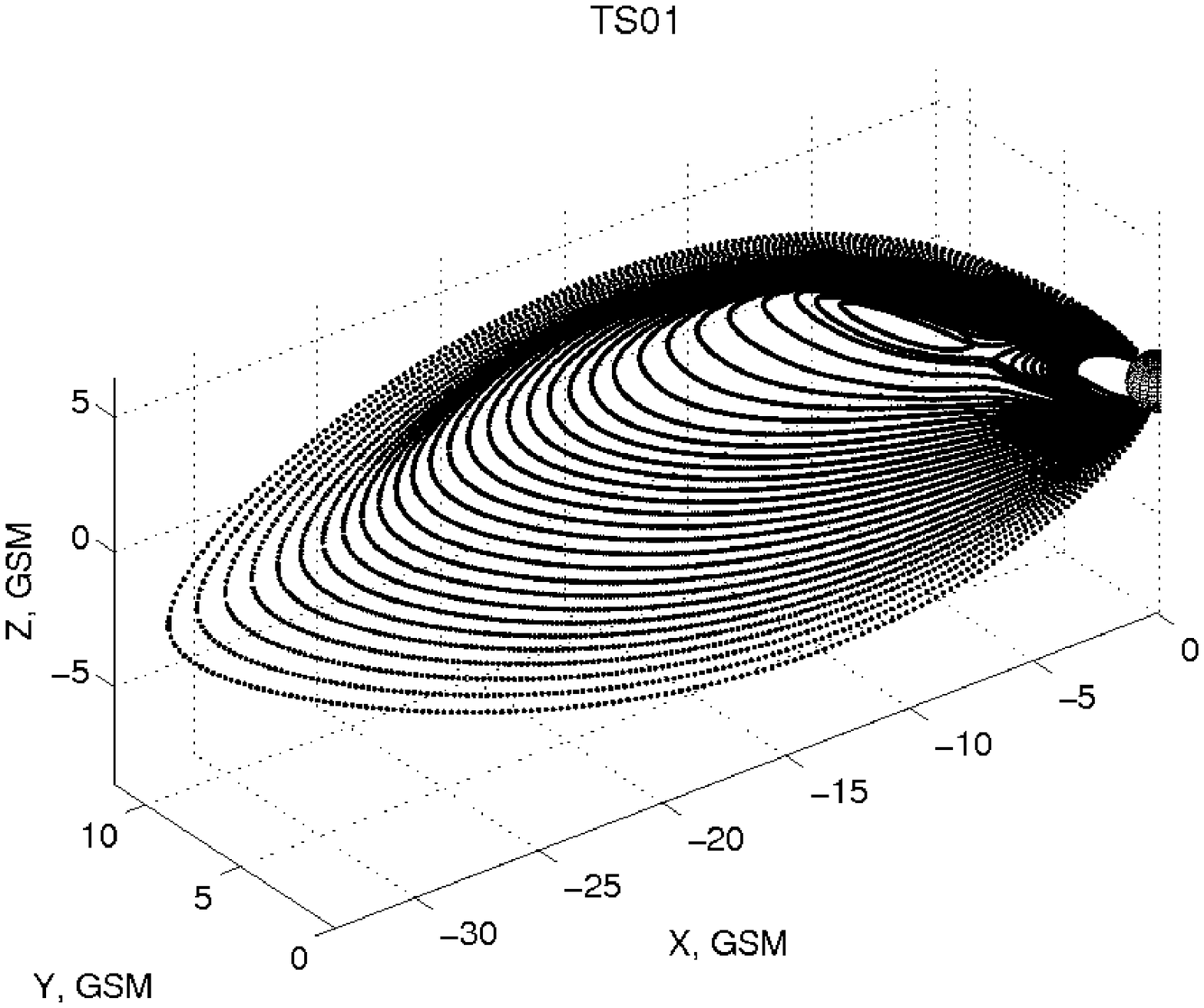}
\label{one}
\end{figure}
\begin{figure}
\centering \vspace{-0.0cm} \hspace{-0.3 cm}
\includegraphics[width=.5\textwidth]{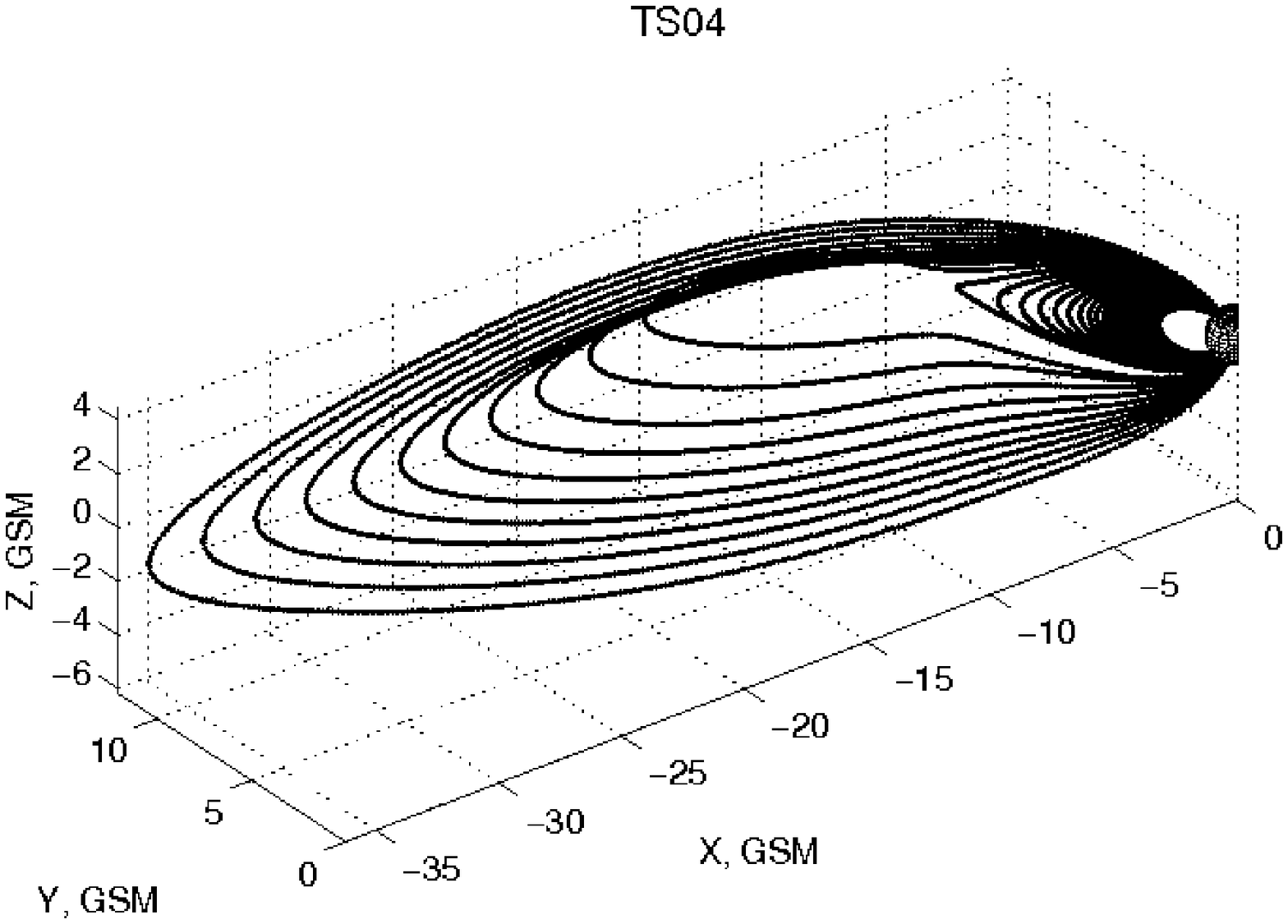}
\caption{The geomagnetic field lines, obtained using the IGRF(a), Tsyganenko 2001(b), and Tsyganenko 2004 (c) geomagnetic field models.}
\label{one}
\end{figure}
The main difficulty in the reconstruction of the low energy plasma pressure profile is how to map the pressure obtained into the equatorial plane. Figure 3 shows the geomagnetic field lines, obtained using the International Geomagnetic Reference Field (IGRF), Tsyganenko 2001 and 2004 geomagnetic field models (Macmillan et al. (2003), Tsyganenko 2002ab, 2005). IGRF model gives a standard mathematical description of the Earth s main magnetic field and does not consider their distortion caused by magnetospheric current systems. Tsyganenko 2001 (TS01) and 2004 (TS04) models represent the expected statistical response of the geomagnetic field to the orientation of the Earth s dipole axis, solar wind pressure, interplanetary magnetic field, and appropriate geophysical indices. TS04 model is a dynamical model of the storm-time geomagnetic field in the inner magnetosphere that includes the temporal evolution of magnetospheric current systems during the entire storm cycle. The input parameters are similar to those used by the TS01 model. Nevertheless a combination of the solar wind density, speed, and magnitude of the southward component of the IMF, averaged every five minutes, makes it possible to reproduce their evolution during great geomagnetic storms.

\begin{figure}[t]
\centering \vspace{-0.0cm} \hspace{-0.3 cm}
\includegraphics[width=.5\textwidth]{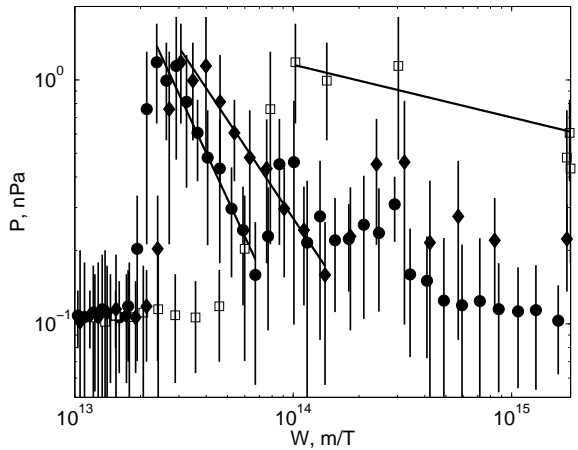}
\caption{Relationship between the plasma pressure, obtained from electron precipitating fluxes, measured by the Aureol-3 satellite and is the flux tube volume per unit magnetic flux obtained using the IGRF (cycles), Tsyganenko 2001(squares), and Tsyganeneko 2004 (diamonds) geomagnetic field models.}
\label{one}
\end{figure}

As it can be seen in Figure 3, the geomagnetic field lines differ significantly for all models, especially for the distances more than 4 Earth s Radii (Re). The TS01 model also has a singularity close to 6 Re in the tail direction. The same picture was observed for other two events, occurred at the end of main phase and in the beginning of recovery phases. For other events the singularities were observed at the distances more than 10 Re.  Therefore, we analyzed the distribution of plasma pressure and other related parameters for the distances closer than the singularities often observed in TS01 and/or TS04 models. 

Stability of plasma pressure profile formed with respect to the development of interchange or flute instability can be an important factor that determines the processes during magnetic storms (see Antonova (2005, 2006) and references therein). It supposes, that increase of the plasma pressure inside the magnetosphere in the region of low plasma parameter $\beta$ ($\beta=2\mu_{0}p/B_{2}$),
where $B$ is the value of the magnetic field in the equatorial plane, ?0 is the permeability of vacuum) by radial plasma transport and flux tube filling by ionospheric source is possible up to the limit determined by the interchange or flute instability development only. According to Kadomtsev (1963), the pressure profile should be lower that a critical limit given by  . Here $W=\int dl/B$ is the flux tube volume per unit magnetic flux, $dl$ is the element of field line length, B is the value of the magnetic field  and integration is done between conjugate hemispheres, $\gamma$ is the ratio of specific heat and the value of const is determined by the boundary conditions. In case of adiabatic compression $\gamma = 5/3$. Particle transport together with conservation of the number of particles in the magnetic flux tube and first and second adiabatic invariants leads to nearly the same value $\gamma = 7/4$ (Tverskoy, 1997).

The analysis of the stability of plasma pressure distribution in such a case requires the analysis of a relationship between the plasma pressure and the magnetic flux tube volume per unit flux $P=P(W)$ (see Figure 4).   The values of $W$ were obtained by integrating the values of geomagnetic field along the field lines using the IGRF (circles), TS01 (squares) and TS04 (diamonds) models.  As can be seen, there is a strong difference in the behavior of plasma pressure according to the IGRF and TS04 from one side and TS01 from other side. For the first two models plasma pressure profile is situated at the dipole-like field lines, whereas TS01 has only three points with negative pressure gradient at the dipole-like field lines. So we included 4 points from the tail-like configuration after singularity to be able to fit the pressure profile, as illustration. Nevertheless, we believe, that change from dipole-like to tail-like configuration should be reflected in particle precipitations as well, what did not happen, indicating that the TS01 geomagnetic field configuration is much less probable. Radial pressure profiles were fitted by a power law $P\left( W\right) W^{\gamma_{W} }$.  Such fittings give the following values for the exponents  $\gamma_{W}=-1.9\pm0.1$ for IGRF,  $\gamma_{W} =-0.21\pm0.08$ for TS01, and  $\gamma_{W} =-1.34\pm0.07$ for TS04. It is easy to see, that the TS01 model is much more overstretched than the IGRF and TS04 models. Nevertheless, it is reasonable to assume that at the distances less that 4Re, the difference between configuration of geomagnetic field lines and dipolar ones is not strong, and the correct value of  $\gamma$ lies between -1.3 and -1.9.  Unfortunately this accuracy is not enough to suggest specific plasma instability, which could be an important factor for the geomagnetic storm development. Nevertheless, it allows tracking the relative changes in the plasma pressure profiles, related to the storm development. 
\begin{figure}
\centering \vspace{-0.0cm} \hspace{-0.3 cm}
\includegraphics[width=.5\textwidth]{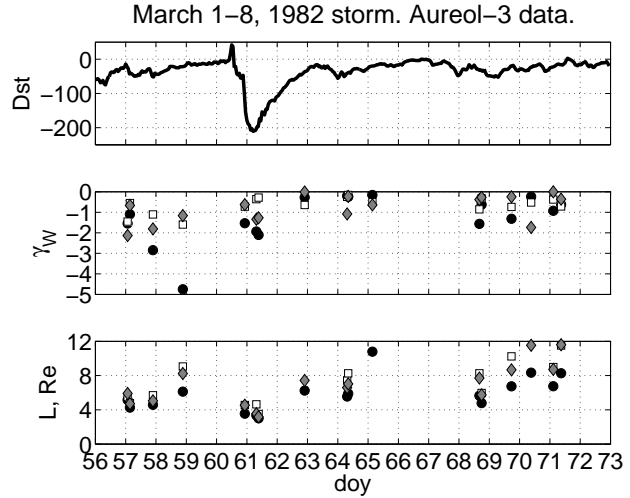}
\caption{From top to bottom: variation of Dst index, variation of the $\gamma_{W}$, obtained due to fitting of $P\left( W\right) W^{\gamma_{W} }$, maximum pressure position in Re. Here black circles were used for IGRF, white squares for TS01, and gray diamonds for TS04 models}
\label{one}
\end{figure}
Figure 5 shows the variation in the exponents  $\gamma_{W}$ for IGRF (black circles), TS01 (white squares), and TS04 (gray diamonds). As it can be seen, during February 26 and 27 (days of year 57 and 58) IGRF model displays strong decrease reaching unrealistic values of  $\gamma_{W}=-4.6$. For this orbit, there are significant differences between IGRF from one side and TS01 and TS04 models from other side. This indicates that for this specific orbit (February 27, 1982 at 22 UT, approximately) the geomagnetic field lines probably deviated on dipolar configuration even at the distances of 6 Re. During the recovery phase all three models indicate that the radial pressure profiles start to be flatter. All three models also indicate that at the end of main phase the maximum of plasma pressure was situated close to the Earth ($L_{max}= 3.0 Re$ for IGRF, Lmax=3.5 Re for TS01,  $L_{max}=3.2 Re$ for TS04). It is interesting to mention that according to Tverskaya (1986), Tverskaya et al. (2003) lowest position of the westward electrojet center Lmax during the storm is related to the maximum value of Dst-variation $Dst_{max}$ in accordance with the relation:
\begin{equation}
D_{st}=2.75 10^{4} L_{max}^{-4} nT
\end{equation}
where $_{Lmax}$  corresponds to the position of the plasma pressure maximum. The relationship (1) was explained by Tverskoy (1997), Antonova (2005, 2006) including the coefficient $2.75 10^{4}$. For the storm analyzed $D_{st}=211$ nT, that corresponds to the theoretical value $L_{max}=3.34 Re$. This value is in a rather good agreement with our estimation of the position of plasma pressure maximum despite we can analyze only part (may be only small part) of integral plasma pressure, excluding the contribution of more energetic particles. 
\begin{figure}
\centering \vspace{-0.0cm} \hspace{-0.3 cm}
\includegraphics[width=.5\textwidth]{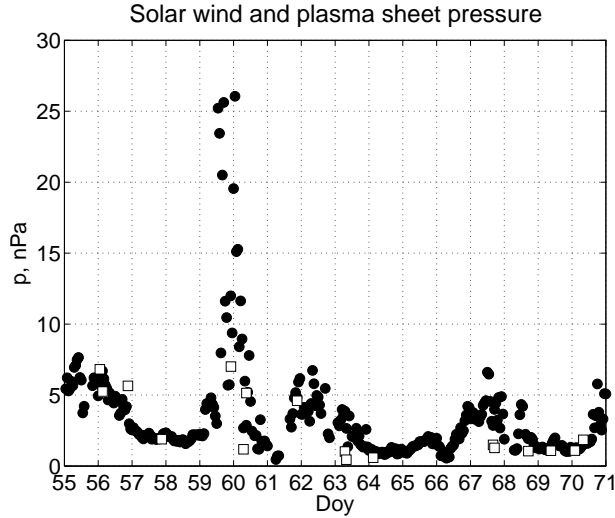}
\caption{Variation of the solar wind dynamic pressure from OMNI2 data base (black circles) and variation of maximum plasma pressure in the inner magnetosphere (white squares).}
\label{one}
\end{figure}
Interesting feature of the observed values of plasma pressure maxima appears when we compare these values with the solar wind dynamic pressure (Figure 6). It is possible to see a very good correlation, between the values of pressure maximum and solar wind dynamic pressure, excluding the main phase of magnetic bay, when the solar wind dynamic pressure is five times larger. We assume that this result must be tested in wider data sets, especially considering great uncertainty of our methodology for obtaining values of plasma pressure maxima inside the magnetosphere. This result agrees with the results of Borovsky at al. (1998) who established that the density of the plasma sheet is strongly correlated with the density of the solar wind.

Correlation between the solar wind dynamic pressure and the maximum plasma pressure in the inner magnetosphere can be a consequence of magnetopause pressure balance during at least quiet periods. Solar wind dynamic pressure is balanced by the pressure of the inner magnetospheric magnetic field only in the regions of low plasma $\beta$ parameter inside the magnetosphere (for example, in the subsolar point). Nevertheless, the plasma $\beta$  parameter can be higher in the near cusp regions and at the magnetospheric flanks. Therefore, contribution of the inner magnetospheric plasma pressure in the total pressure balance in these regions can be very high. Taking into consideration, that in case of magnetostatic equilibrium the plasma pressure is constant along current lines, it is possible to conclude that quiet time current lines that correspond to the region of plasma pressure maximum cross the magnetopause or close itself inside the magnetosphere in the near cusp regions, where plasma pressure must be close to the solar wind dynamic pressure. 

This result can be important for understanding the processes of the solar wind-magnetosphere coupling. Recently,  Newell at al. (2006, 2007) proposed a "universal solar-wind magnetosphere coupling function" that represents the rate of change of the magnetic flux at the magnetopause and depends on the solar wind velocity, the value of interplanetary magnetic field, and the IMF clock angle: $dF_{MF}/dt=v^{4/3}B_{T}^{2/3}sin ^{8/3}\left( \alpha/2\right)$ . The function proposed gives the best fits for all tested ground based and satellite indices, except the Dst. Nevertheless, the authors argued "that using either the current value of $p ^{1/2}$ or using $p ^{1/2}$ integrated over 72 hours improved Dst. ... this implies that only a portion of the advantage of "correcting" Dst for $p ^{1/2}$ actually arises because the ring current perturbation on the ground-station magnetometers. In other words, the actual ring current itself does respond to  $p ^{1/2}$, presumably because of the increased proximity of the magnetopause". Our results indicate about the possibility of a direct influence of the solar wind dynamic pressure on the pressure in the inner magnetosphere. 

\section{Discussion and Conclusions}

In this paper we present an analysis of the distribution of lower energy part of plasma pressure inside the magnetosphere and reveal some characteristic features of plasma pressure evolution during geomagnetic storm.  In particular, we found that the plasma pressure profile became steeper,  especially near the end of the main phase, having the value near and even exceeding (in IGRF mapping) the maximum steeping value $\gamma_{W}$=7/4 necessary for the development of interchange instability in the case of isotropic pressure and near to dipole magnetic configuration. 
Unfortunately empirical storm time models of the magnetospheric magnetic field are not well developed. This fact is reflected in significant differences in the plasma pressure profiles, obtained using Tsyganenko 2001 and 2004 geomagnetic field models, that demands additional studies of inner magnetosphere pressure distribution using a self-consistent  magnetospheric model in which the magnetic forces are equilibrated by plasma pressure forces, satisfying a condition of quasi-static equilibrium (such equilibrium is realized when plasma velocity is much smaller than Alfven and sound velocities). One of the first important steps in such direction was made by Zaharia et al., (2005, 2006). 
Storm time plasma pressure profiles at L<9 observed by AMPTE/SSE satellite (Lui et al., 1987) give the values of plasma parameter $\beta<1$.  In our case all three used models positioned the maximum of plasma pressure always at quasi-dipole magnetic field lines, where the magnetic pressure was hundreds of times higher than the plasma pressure. We measure only a low energy portion of plasma pressure. Nevertheless, there still exists a probability of high  plasma in the region of plasma pressure maximum. However, taking into account the modeling results of Zaharia et al. (2005, 2006) it is possible to suggest that low  approximation is valid at least for analyzed magnetic storm.

It is interesting to mention that it was also found that the position of pressure maximum follows well the experimentally obtained dependence of the position of auroral electrojet on the value of Dst variation. This coincidence provides additional support of the theory of Dst formation during magnetic storms developed by Tverskoy (1987), Antonova (2005, 2006).

Interesting feature appears when we compare the values plasma pressure maxima with the solar wind dynamic pressure. A very good correlation is observed for all satellite passages, excluding the main phase of magnetic bay, when the solar wind dynamic pressure is five times larger. We plan to study this effect in detail in the nearest future.

\textbf{Acknowledgments}
The authors acknowledge the Centre National d'Etudes Spatiales, France, and especially to Mrs. J. Cuvilo (CESR) and C. Guerin (CETP) for creating the Aureol-3 database. We thank WDC-C2 Kyoto AE index service, AE stations and the people who derive the index. We acknowledge A. Lazarus and MIT Space Plasma Group for providing the IMF data, and A. Szabo and R. Lepping (NASA GSFC) for providing the plasma data. We also acknowledge the NSSDC for providing the IRGF, Tsyganenko 2001 and 2004 geomagnetic field models. This work was made possible by the support of the Department of Research in Science and Technology (DICYT), Chilean National Foundation for Science and Technology (FONDECYT) grant 1070131, Chilean Air Force Aeronautic Polytechnic Academy, and the RFBR grant 05-05-64394-a (Russia). 

\textbf{REFERENCES}

Antonova, E.E. Storm-substorm relations and high latitude currents. 
{\it Adv. Space Res., 30(10) }, 2219-2224, 2002.
 
Antonova, E.E. Magnetostatic equilibrium and current systems in the Earth's magnetosphere, 
{\it Adv. Space Res., 33}, 752-760, 2004. 

Antonova E.E. Magnetospheric substorms and the sources of inner magnetosphere particle acceleration, The Inner Magnetosphere: Physics and Modeling, {\it Geophysical Monograph Series 155}, Copyright 2005 by AGU, doi:10.1029/155GM12, p. 105-111, 2005. 

Antonova E.E. Stability of the magnetospheric plasma pressure distribution and magnetospheric storms, {\it Adv. Space Res., 38(8)}, 1626-1630, 2006.

Antonova, E.E., and Ganushkina, N. Yu.  Inner magnetospheric currents and their role in the magnetosphere dynamics, {\it Phys. Chem. Earth(C), 25(1-2)}, 23-26, 2000.  

Antonova, E. E., and B. A. Tverskoy, On the nature of inverted-V electron precipitation band and Harang discontinuity in the evening sector of auroral ionosphere, {\it Geomagn. Aeron., Engl. Transl., 15(1),} 105-111, 1975.

Borovsky, J. E., Thomsen, M. F.; Elphic, R. C., The driving of the plasma sheet by the solar wind, 	{\it J. Geophys. Res., 103(A8),} 17617-17640, 1998.

Bosqued, J. M., Maurel, C., Sauvaud, J. A., Kovrazhkin, R. A., Galperin, Yu. I. Observations of auroral electron inverted-V structures by the AUREOL-3 satellite. {\it Planet. Space Sci., 34 (3)}, 255-269, 1986.

Daglis, I. A. The role of magnetosphere-ionosphere coupling in magnetic storm dynamics. Magnetic storms, {\it Geophysical Monograph 98}, 107-115, 1997. 

DeMichelis, P., Daglis, I. A.,  Consolini, G.  An average image of proton plasma pressure and of current systems in the equatorial plane derived from AMPTE/CCE-CHEM measurements, {\it J. Geophys. Res., 104(A12)}, 28615-28624, 1999.  

Dessler, A.J.,  and Parker, E.N.  Hydromagnetic theory of geomagnetic storms, {\it J. Geophys. Res., 64(12)}, 2239-2252, 1959. 

Iyemori, T., Rao, D.R.K. Decay of the Dst field of geomagnetic disturbance after substorm onset and its implications to storm-substorm relation, {\it Ann. Geophys., 14(6)}, 608-618, 1996. 

Kadomtsev, B.B. Hydrodynamic stability of plasma, in: Leontovich, M.A. (Ed.), {\it Problems of Plasma Theory}, vol. 2. pp. 132-176, 1963.

Kozelova, T.V., Treihou, J.P., Korth, A., Kremser, G., Lazutin, L.,  Melnikov, A.O., Pedersen, A., Sakharov, Y.A., Substorm active phase study by ground-based and satellite measurements. {\it Geomagn. Aeron., 26(6)}, 963-969, 1986.  

Liemohn, M.W. Yet another caveat to using Dessler-Parker-Scopke relation, {\it J. Geophys. Res., 108 (A6)}. doi:10.1029/2003JA009839, 2003.

Lui, A.T.Y., McEntire, R.W.,  Crimigis, S.M.  Evolution of the ring current during two geomagnetic storms, {\it J. Geophys. Res., 92(A7),}  7459-7470, 1987.

Lui, A.T.Y., Hamilton, D.C. Radial profile of quite time magnetospheric parameters, {\it J. Geophys. Res., 97(A12)}, 19,325-19,332, 1992. 

Lui, A.T.Y., Inner magnetosphere plasma pressure distribution and its local time asymmetry, {\it Geophys. Res. Lett., 30(16),} 1846, doi:10,1029/2003GL017596, 2003.

Lyons, R.L., Williams, D.J. Quantitative aspects of magnetospheric physics. D Reidel Publ. Co., Dordrecht, Boston, Lancaster, 1982, 231 p.

Lyons, R.L., Wang, C.-P., Nagai, T., Mukai, T., Saito, Y., Samson, J.C. Substorm inner plasma sheet particle reduction, {\it J. Geophys. Res., 108(A12)}, doi:10.1029/2003JA010177, 2003. 

Macmillan, S., Maus, S., Bondar, T., Chambodut, A., Golovkov, V., Holme R., Langlais, B., Lesur, V., Lowes, F., Luhr, H., Mai, W., Mandea, M., Olsen, N., Rother, M., Sabaka, T., Thomson, A., Wardinski, I., Ninth Generation International Geomagnetic Reference Field Released, EOS Transactions, AGU, Vol. 84, Issue 46, 503, 2003.

McPherron, R.L. The role of substorms in the generation of magnetic storms. Magnetic storms, {\it Geophysical Monograph 98}, 131-147, 1997. 

Newell, P. T., Sotirelis, T., Liou, K., Meng, C.-I. and Rich F. J., Cusp latitude and the optimal solar wind coupling function, {\it J. Geophys. Res., 111, A09207}, doi:10.1029/2006JA011731, 2006.

Newell P. T., T. Sotirelis, K. Liou, C.-I. Meng, F. J. Rich, A nearly universal solar wind-magnetosphere coupling function inferred from 10 magnetospheric state variables, {\it J. Geophys. Res., 112, A01206}, doi:10.1029/2006JA012015, 2007.

Parker, E.N. The alternative paradigm for magnetospheric physics. {\it J. Geophys. Res., 101(A5),} 10587-10625, 1996. 

Pulkkinen, T.I., Ganushkina, N. Yu., Baker, G.N., Turner, N.E., Fennell, J.F., Roeder, J., Fritz, T.A., Grande, M., Kellett, B., Kettmann, G. Ring current ion composition during solar minimum and rising solar activity: Polar/CAMMICE/MICS results. {\it J. Geophys. Res. 106(A9)}, 19131-19147, 2001. 

Roux, A.S. Generation of field-aligned current system at substorm onset, Proc. {\it ESA Workshop on Future missions in solar, heliospheric and space plasma physics}, Germany, Garmisch-Parten Kirchen, Germany 30 April-3 May, ESA SP-235, 151-159, 1985.  

Sckopke, N., A general relation between energy of trapped particles and the disturbance field over the Earth. {\it J. Geophys. Res., 71(13)}, .3125-3130, 1966. 

Stepanova M.V., Antonova, E.E.,  Bosqued, J.M., Kovrazhkin, R.A., Aubel, K.R. Asymmetry of auroral electron precipitations and its relationship to the substorm expansion phase onset,{\it  J. Geophys. Res., 107(A7)}, doi:10.1029/2001JA003503, 2002.

Stepanova, M.V., Antonova, E.E., Bosqued, J.M., Kovrazhkin, R.A. Azimuthal plasma pressure reconstructed by using the Aureol-3 satellite data during quiet geomagnetic conditions. {\it Adv.  Space Res., 33(5)}, 737-741, doi:10.1016/S0273-1177(03)00641-0, 2004a. 

Stepanova, M., Antonova, E. E., J. Bosqued, J. M., Kovrazhkin, R. A. Radial plasma pressure gradients in the high-latitude magnetosphere as sources of instabilities leading to the substorm onset, {\it Advances in Space Research, 33(5)}, 761-768, doi:10.1016/S0273-1177(03)00634-0, 2004.

Sugiura, M., Hourly values of equatorial Dst for the IGY, {\it Ann. Int. Geophys. Year, 35,} 9, Pergamon Press, Oxford, 1964. 

Tverskoy, B.A. Formation mechanism for the structure of the magnetic storm ring current. {\it Geomagn. Aeron., 37(5)}, 555-559, 1997.

Tverskaya. L.V. On the boundary of electron injection into the magnetosphere, {\it Geomagn. Aeron., 26(5)}, 864-865, 1986.  

Tverskaya, L.V., Pavlov, N.N., Blake, J.B., Selesnick, R.S., Fennell, J.F. Predicting the L-position of the storm-injected relativistic electron belt, {\it Adv. Space Res., 31(4)}, 1039-1044, 2003. 

Tsyganenko, N.A. A model of the near magnetosphere with a dawn-dusk asymmetry - 1. Mathematical Structure. {\it J. Geophys. Res., 107(A8)}, doi:10.1029/2001JA000219, 2002a.  

Tsyganenko, N.A.  A model of the near magnetosphere with a dawn-dusk asymmetry - 2. Parameterization and fitting to observations. {\it J. Geophys. Res., 107(A7),} doi:10.1029/2001JA000220, 2002b.  

Tsyganenko, N. A.,  Sitnov, M.I. Modeling the dynamics of the inner magnetosphere during strong geomagnetic storms. {\it J. Geophys. Res., 110(A3)}, doi:10.1029/2004JA010798, 2005.  

Williams,  D. J., The Earth's ring current: Causes, generation and decay, {\it Space Sci. Rev., 34(1/2)}, 223-234, 1983.

Wing, S., Newell, P. T. Central plasma sheet ion properties  as inferred from ionospheric observations. {\it J. Geophys. Res., 103(A4)}, 6785-6800, 1998. 

Wing, S., Newell, P. T. Quite time plasma sheet ion pressure contribution to Birkeland currents, {\it J. Geophys. Res., 105(A4)}, 7793-7802, 2000. 

Zaharia, S., J. Birn, C. Z. Cheng, Toward a global magnetospheric equilibrium model, {\it J. Geophys. Res., 110(A9)}, doi: 10.1029/2005JA011101, 2005.

Zaharia, S., Jordanova, V. K., Thomsen, M. F., Reeves, G. D. Self-consistent modeling of magnetic fields and plasmas in the inner magnetosphere: Application to a geomagnetic storm, {\it J. Geophys. Res., 111, A11S14}, doi:10.1029/2006JA011619, 2006.

\end{document}